\documentclass[
showpacs,preprintnumbers,amsmath,amssymb,floatfix]{revtex4}

\usepackage{graphicx}
\usepackage{dcolumn}
\usepackage{bm}

\usepackage{amsmath}
\usepackage{psfrag}
\usepackage{color}
\usepackage{latexsym,amsfonts}
\usepackage{graphicx}
\providecommand{\href}[2]{#2}   

\def\conj#1{\buildrel{*}\over{#1}}
\def\la{\lambda}
\def\La{\Lambda}
\def\ve{\varepsilon}
\def\<{\langle}
\def\>{\rangle}

\def\half{{^1\!/_2}}

\def\ch{\mathop{\rm ch}\nolimits}
\def\sh{\mathop{\rm sh}\nolimits}

\def\n{h}
\def\a{\theta}
\def\Wth{W_{\rm th}}
\def\Wmax{W_{\rm max}}
\def\deltaPk{\delta(P\!+\!k\!-\!P'')}

\begin{document}


\title{Beam normal spin asymmetry of elastic $eN$ scattering in the leading logarithm approximation}
\author{Dmitry~Borisyuk}
\author{Alexander~Kobushkin}%
\email{kobushkin@bitp.kiev.ua}
\affiliation{Bogolyubov Institute for Theoretical Physics\\
Metrologicheskaya str. 14-B, 03143, Kiev, Ukraine}

\date{\today}

\begin{abstract}
The beam normal spin asymmetry for the elastic $eN$ scattering is studied in the leading logarithm approximation. We derive the expression for the asymmetry, which is valid for any scattering angles. The result is compared with the results of other authors, obtained for the forward kinematics. We also calculate the numerical values of the asymmetry at intermediate energy and show that they are consistent with existing experimental data.
\end{abstract}

\maketitle

\section{Introduction\label{sec:Introduction}}
During last few years a lot of effort has been made in experiment and theory to get an information about effects beyond the one photon exchange approximation in the elastic $eN$ scattering. Such a study is important to get reliable information on the nucleon structure at short distances.

In light of this, one-particle polarization observables, the target and beam normal spin asymmetry, are of special interest due to the following circumstance. First, these quantities disappear in the one-photon approximation and thus give a direct information on the two-photon exchange in the $eN$ scattering. Second, they are proportional to imaginary part of the amplitude, which is easier for theoretical analysis, than the real part in the second order of the perturbation theory. 

Recently first experimental results on the beam normal asymmetry in the elastic $ep$ scattering, $B_n$, have been reported \cite{SAMPLE,MAMI}.

Theoretical analysis of $B_n$ was done in the framework of different models \cite{AfanAkushMer,DiakR-M,GGV,Pasq,AfanMer,Gorchtein}. The asymmetry $B_n$ is proportional to the fine structure constant $\alpha\approx1/137$, as well as to the electron mass $m$. Important feature of this quantity is that it contains terms, proportional to the large logarithms, $\ln^2(Q^2/m^2)$ and $\ln(Q^2/m^2)$, $Q^2=-q^2$, where $q^2$ is the momentum transfer squared.

The latter effect was discussed for the forward kinematics at high electron energy \cite{AfanMer,Gorchtein}. However somewhat different approach was used in \cite{AfanMer}, see Sec.~\ref{sec:Forward} for the detailed consideration.

In the present paper we calculate the leading logarithmic contribution to $B_n$ at any momentum transfer and energy. For the forward kinematics we obtain the $Q$-dependence identical to \cite{Gorchtein}, nevertheless, the beam asymmetry has different relation with electromagnetic transition amplitudes.  

The paper is organized as follows. In Section~\ref{sec:Kinematics} we discuss kinematics and introduce notation. Basic formulae and the general structure of the beam normal spin asymmetry in terms of hadronic and leptonic tensors are given in Section~\ref{sec:General}. The double-logarithmic contribution is calculated in Section~\ref{sec:DLogs}. In Section~\ref{sec:Forward} we consider the forward kinematics and compare our results with the existing calculations. Numerical results and comparison with experiment are given in Section~\ref{sec:Numerical}. Concluding remarks are summarized in Section~\ref{sec:Conclusions}.
\section{Kinematics and notation\label{sec:Kinematics}}
We consider the reaction
\begin{equation}
 e + N \to e + N.
\end{equation}
The initial electron and nucleon momenta are denoted $k$ and $P$, respectively,
and the final momenta $k'$ and $P'$. The transferred momentum is $q=k-k'$ ($q^2<0$), and the c.m. energy squared is $s=(P+k)^2=(P'+k')^2$. Nucleon and electron masses are denoted $M$ and $m$, respectively. We will $not$ neglect the electron mass in the following.

\begin{figure}[t]
\centering
\psfrag{k}{$k$}\psfrag{k1}{$k'$}\psfrag{k2}{$k''$}
\psfrag{q1}{$q_1$}\psfrag{q2}{$q_2$}
\psfrag{p1}{$P$}
\psfrag{p2}{$P'$}
\psfrag{p3}{$P''$}
\includegraphics[width=0.35\textwidth]{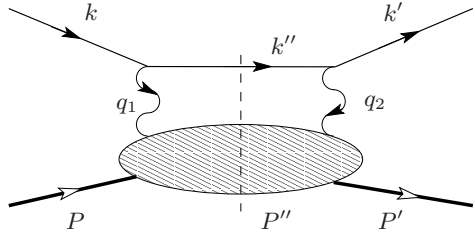}
\caption{Two photon exchange for the elastic $eN$ scattering.}
\label{Twophot}
\end{figure}

The target nucleon is nonpolarized, the beam electron is polarized along the normal to the reaction plane. There are two possibilities: its spin 4-vector can be either $S_\mu$ or $-S_\mu$, where
\begin{equation}
  S_\mu = A \ve^{\nu\mu\sigma\tau} k_\nu P_\sigma P'_\tau, \quad A>0 {\rm\ such\ that\ } S^2=-1.
\end{equation}
Denoting corresponding cross-sections $\sigma_\uparrow$ and $\sigma_\downarrow$,
we define the beam normal asymmetry by
\begin{equation}
 B_n = {\sigma_\uparrow-\sigma_\downarrow \over \sigma_\uparrow+\sigma_\downarrow}.
\end{equation}
It can be shown that $B_n$ is proportional to the absorptive part of the
scattering amplitude and 
unitarity can be used to evaluate it.
In the lowest order of $\alpha$ the intermediate state, entering the unitarity
relation, consists of the electron (with momentum $k''$) and some hadronic state; its momentum is denoted by $P''=P+q_1=P'-q_2$, see Figure~\ref{Twophot}. The mass of the hadronic intermediate state is $W=\sqrt{P''^2}$. For inelastic ($\neq$ proton) states it varies from $\Wth = M+m_\pi$ (where $m_\pi$ is pion mass) to its maximal possible value $\Wmax = \sqrt{s} - m$.

In the c.m. frame we use the following notations for components of the electron momenta $k=(\epsilon,\vec k)$, $k'=(\epsilon,\vec k')$ and $k''=(\epsilon'',\vec k'')$. The incoming (outgoing) electron energy $\epsilon$ and the intermediate electron energy $\epsilon''$ read
\begin{equation}
  \epsilon = {s-M^2+m^2 \over 2\sqrt{s}} \;,\qquad \epsilon'' = {s-W^2+m^2 \over 2\sqrt{s}}\;.
\end{equation}
The scattering angle (between $\vec k$ and $\vec k'$) is $\a$, the angles between $\vec k''$ and the momenta $\vec k$ and $\vec k'$ are $\a_1$ and $\a_2$, respectively. The transferred momenta squared are
\begin{equation}
q^2 = -Q^2 = -4 |\vec k|^2 \sin^2 \a/2, \quad
q_1^2 = \Delta - 4|\vec k||\vec k''|\sin^2 \a_1/2,
\quad
q_2^2 = \Delta - 4|\vec k||\vec k''|\sin^2 {\a_2/2},
\end{equation}
where $\Delta = -2(\epsilon\epsilon''-|\vec k||\vec k''| - m^2) < 0$.

We denote Dirac matrices $\gamma_\mu$ and $\hat a \equiv a_\mu\gamma^\mu$.

Electromagnetic current matrix elements for the nucleon are
\begin{equation} \label{Jproton}
\<P' \la' |J_\mu|P \la \> = \<P'\la'| \Gamma_\mu |P\la\>
=\<P'\la'| \left[ 2M(G_E-G_M){P^\mu_+ \over P_+^2} + G_M \gamma^\mu \right]
 |P\la\>,
\end{equation}
where $|P\la\>$ is the nucleon state with momentum $P$ and helicity $\la$,
$P_+ = P+P'$, $G_E \equiv G_E(q^2)$ and $G_M \equiv G_M(q^2)$
are nucleon elastic formfactors.

Current matrix elements between the nucleon and other hadronic states
can be expressed via 3 independent invariant amplitudes $f_\la$ with $\la=\pm 1,0$:
\begin{equation} \label{efs}
\ve^{(\la)}_\mu \<\n \La |J^\mu| P \half \> =
  f^{(\n)}_\la(q^2) \delta_{\La,\la + \half}, \quad
\ve^{(-\la)}_\mu \<\n -\!\La |J^\mu| P -\!\half \> =
  \eta_\n f^{(\n)}_\la(q^2) \delta_{\La,\la + \half}.
\end{equation}
Here $\eta_\n = \pi_\n e^{i \pi(s_\n-\half)}$, $\n$ is the intermediate hadronic state, $s_\n,\pi_\n$ are its spin and parity, $\La$ is spin projection onto the vector $\vec P$. The quantities $f_\la^{(\n)}$ can be considered as helicity amplitudes of the process $\gamma^* N \rightarrow \n$. The $\ve_\mu^{(\la)}$ are polarization vectors of virtual photon with helicity $\la$. They satisfy orthogonality relations
\begin{equation}
  \sum_\la (-1)^\la \ve^{(\la)}_\mu \conj{\ve}\!^{(\la)}_\nu
   = g_{\mu\nu} - {q_\mu q_\nu \over q^2}, \quad 
  \ve^{(\la)}_\mu \conj{\ve}\!^{(\la')\mu} =
  (-1)^\la \delta_{\la \la'}.
\end{equation}

\section{Basic formulas\label{sec:General}}
We start from the general formula for $B_n$,
which can be derived, using unitarity, in the same manner as the target
asymmetry in \cite{deRuj},
\begin{equation} \label{asy1}
  B_n ={i \alpha q^2 \over 2 \pi^2 D}
    \int {d^3 k'' \over 2\epsilon''}{1 \over q_1^2 q_2^2} L^{\alpha\mu\nu}\sum_{\la_p,\la'_p}
    W_{\mu\nu}(P'\la'_p;P\la_p)\ \<P\la_p| \Gamma_\alpha |P'\la'_p\>,
\end{equation}
where
\begin{equation}
  D = 4 \left[ { (2s - 2M^2 - Q^2)^2 \over 4 M^2 + Q^2}
  (4 M^2 G_E^2 + Q^2 G_M^2) - Q^2 (4 M^2 G_E^2 - Q^2 G_M^2) \right],
\end{equation}
the hadronic tensor $W_{\mu\nu}$ is defined as
\begin{equation}
  W_{\mu\nu}(P'\la'_p;P\la_p) = \sum_\n (2\pi)^4
  \delta(P\!+\!k\!-\!P''\!-\!k'') \<P'\la'_p|J_\mu|\n\> \<\n|J_\nu|P\la_p\>
\end{equation}
and the leptonic tensor is
\begin{equation}
  L^{\alpha\mu\nu} =
  - {\rm Tr\;} (\hat k' + m) \gamma^\mu (\hat k'' + m) \gamma^\nu (\hat k + m)
  \gamma^5 \hat S \gamma^\alpha.
\end{equation}

Contrary to the target normal asymmetry $A_n$, the asymmetry $B_n$ vanishes in the $m \to 0$ limit.
It can be easily seen from the expression for $L^{\alpha\mu\nu}$.
Thus  we cannot neglect $m$ completely, but instead we will systematically
neglect $o(m)$ terms. Another significant feature of $B_n$ is that it contains
logarithmic and double-logarithmic terms $\sim m \ln{Q^2 \over m^2}$ and
$\sim m \ln^2 {Q^2 \over m^2}$, which is not the case for $A_n$.

Consider the integral in (\ref{asy1}) written in general form as
\begin{equation}
  I = m \int {d^3 k'' \over 2\epsilon''}{1 \over q_1^2 q_2^2} Y(W,q_1^2,q_2^2) + o(m).
\end{equation}
There is no explicit $m$ dependence in $Y$, but the indirect dependence
(through $q_1^2$ and $q_2^2$) is present. The function $Y$ can be assumed to
be symmetric under the exchange of $q_1^2$ and $q_2^2$.
If we put $m=0$ in the integrand, the integral will have two types of singularities. The first singularity appears when $q_1^2 \to 0$, but $q_2^2$ is finite or vice versa, $q_2^2 \to 0$, but $q_1^2$ is finite. The second singularity appears when $W \to W_{\rm max}$; in this case both $q_1^2$ and $q_2^2 \to 0$. However, for $m \neq 0$ the integral is nonsingular; those ``singularities'' result in the above-mentioned $\ln m$ and $\ln^2 m$ terms. To isolate the singularities, we rewrite the function $Y$ as
\begin{equation}
  Y(W,q_1^2,q_2^2)  =\widetilde Y(W,q_1^2,q_2^2) + [Y_0(W)-Y_0(W_{\rm max})] +Y_0(W_{\rm max}),
\end{equation}
where $Y_0(W) = Y(W,0,-4|\vec k||\vec k''| \sin^2 {\a/2})$,
$\widetilde Y = Y - Y_0$. After that we integrate each addendum separately, neglecting
the terms which are zero in the $m \to 0$ limit (for the details see Appendix). The result is
\begin{equation} \label{int}
\begin{split}
 \int {d^3 k'' \over 2 \epsilon''}& {1 \over q_1^2 q_2^2} Y(W,q_1^2,q_2^2) = 
 \int {d^3 k'' \over 2 \epsilon''} {1 \over q_1^2 q_2^2} \widetilde Y(W,q_1^2,q_2^2) - {2\pi \over Q^2} \int\limits_{\Wth^2}^s
  \ln \left( {Q \over m}{s-W^2 \over W^2-M^2} \right)
  {Y_0(\sqrt{s}) - Y_0(W) \over s-W^2} dW^2 + \\
& + {2\pi \over Q^2} Y_0(\sqrt{s}) \left[
  {1 \over 8} \ln^2 {Q^2 \over m^2} + {\pi^2 \over 8}
  - {1 \over 4} F(-\cos^2 {\a/2})
  - F(\delta) - \ln (1+\delta) \ln {Q \over m \delta} \right],
\end{split}
\end{equation}
where $\delta = {\Wth^2-M^2 \over s-\Wth^2}$ and $F$ is the Spence function,
\begin{equation} \label{Spence}
  F(\xi) = \int\limits_0^\xi {\ln (1+x) \over x} dx.
\end{equation}
All essential $m$ dependence is written out explicitly, otherwise one can put $m=0$. For this reason we write in the above formula, for instance, $\sqrt{s}$ instead of $W_{\rm max}$.

\section{Leading logarithm approximation\label{sec:DLogs}}

Later on we will study the double-logarithmic contribution only, i.e. we put
\begin{equation}
  \int {d^3 k'' \over 2 \epsilon''} {1 \over q_1^2 q_2^2} Y(W,q_1^2,q_2^2)
  \approx {\pi Y_0(\sqrt{s}) \over 4 Q^2} \ln^2 {Q^2 \over m^2}\;.
\end{equation} 
The logarithm is quite large even for relatively small $Q^2$,
e.g. if $Q^2=0.25{\rm\ GeV^2}$, then $\ln^2{Q^2\over m^2} \approx 200$, and
it should give a large, if not dominant, contribution to the full answer.

Now we will evaluate $Y_0(\sqrt{s})$.

The hadronic tensor was calculated in \cite{BK}. The condition $W=\sqrt{s}$
implies that the rest frame of the intermediate hadronic state,
used in \cite{BK}, coincides with the c.m. frame and
$q_1^2=q_2^2=-2m(\epsilon-m) \approx 0$. The answer is
\begin{eqnarray}
  & W_{\mu\nu}(P' \la'_p; P \la_p) =
  \sum\limits_{\la,\la'=\pm 1}
  \ve_{1\nu}^{(2\la_p \la)} \conj{\ve}\!_{2\mu}^{(2 \la'_p \la')} \times \\
  & \times {\sum\limits_\n}' (2\pi)^4 \delta(P+k-P'')
  f^{(\n)}_\la(0) \conj f\!^{(\n)}_{\la'}(0)
  \ \eta_\n^{\la_p- \la'_p}
  {\cal D}_{\la_p(2\la+1),\la'_p(2\la'+1)}^{(s_\n)}(0,\a,0), \nonumber
\end{eqnarray}
where $\cal{D}$ is Wigner D-function and $\sum_{\n}'$ means sum over all intermediate hadronic states, but without summation over total angular momentum projections (for more detail see \cite{BK}).

The leptonic tensor $L_{\alpha\mu\nu}$ can be rewritten as
\begin{equation}
\begin{split}
  L_{\alpha\mu\nu} = - \sum_{\la_e,\la_e',j_z}
  \<k'\la_e'|\gamma_\mu|k''j_z\> \<k''j_z|\gamma_\nu|k\la_e\>
  \<k\la_e|\gamma^5 \hat S \gamma_\alpha|k'\la_e'\>
\equiv -\sum_{\la_e,\la_e'}l_{\mu\nu}(k'\la_e';k\la_e)\<k\la_e|\gamma^5 \hat S \gamma_\alpha|k'\la_e'\>,
\end{split}
\end{equation}
where $\la_e$, $\la_e'$ are electron helicities, $j_z$ is spin projection of the intermediate electron (one cannot use helicity for the electron at rest).
 The tensor $l_{\mu\nu}$ can be calculated similarly to the hadronic tensor $W_{\mu\nu}$; the difference is that (i) the only intermediate state is electron and (ii) the virtual photons, {\it absorbed} by the nucleon, are {\it emitted} by the  electron and therefore their polarization vectors $\ve_{1\mu}$, $\ve_{2\mu}$
should be conjugated. We have
\begin{equation}
  l_{\mu\nu}(k' \la'_e; k \la_e) =
  \sum\limits_{\la,\la'} (-1)^{\la + \la'}
  \conj{\ve}\!_{1\nu}^{(2\la_e \la)} \ve_{2\mu}^{(2 \la'_e \la')}
  f^{(e)}_\la(q_1^2) \conj f\!^{(e)}_{\la'}(q_2^2)
  {\cal D}_{\la_e(2\la+1),\la'_e(2\la'+1)}^{(\half)}(0,\a,0)
\end{equation}
with the following transition amplitudes for the electron 
\begin{equation}
  f_1^{(e)}(q^2) \equiv 0,\quad f_0^{(e)}(q^2) = 2m,\quad
  f_{-1}^{(e)}(q^2) = -\sqrt{-2q^2}.
\end{equation}
Since $q_1^2=q_2^2=-2m(\epsilon-m)$ we see that $f^{(e)}_0/f^{(e)}_{-1} \sim \sqrt{m}$ and $f^{(e)}_0$ is to be neglected. After that
\begin{equation}
  l_{\mu\nu} = 4m\epsilon
    \conj{\ve}\!_{1\nu}^{(-2\la_e)} \ve_{2\mu}^{(-2\la'_e)}
    \cos{\a+\pi(\la_e-\la_e') \over 2} \;.
\end{equation}
Now the lepton tensor becomes
\begin{equation}
  L_{\alpha\mu\nu} = - 4m\epsilon \sum_{\la_e,\la_e'}
    \conj{\ve}\!_{1\nu}^{(-2\la_e)} \ve_{2\mu}^{(-2\la'_e)}
    \cos{\a+\pi(\la_e-\la_e') \over 2}
    \<k\la_e|\gamma^5 \hat S \gamma_\alpha|k'\la_e'\>\;.
\end{equation}
The operator $\gamma^5 \hat S \gamma^\alpha$ flips the electron helicity. Indeed, $\gamma^5 \hat S$ is the operator of spin projection onto the axis, orthogonal to the electron momentum and thus flips the helicity, while $\gamma^\alpha$ is the operator of the electromagnetic current, which is helicity-conserving. So one concludes that $\la_e = -\la_e'$ and the virtual photons should have opposite helicities. This is the reason why $B_n$ cannot be expressed through the total photoabsorption cross-section.

After that the leptonic tensor reads
\begin{equation}
  L_{\alpha\mu\nu} = 4m\epsilon \sin {\a/2} 
    \sum_{\la_e} 2\la_e \conj{\ve}\!_{1\nu}^{(-2\la_e)} \ve_{2\mu}^{(2\la_e)}
    \<k\la_e|\gamma^5 \hat S \gamma_\alpha|k'-\!\la_e\> \;.
\end{equation}

To proceed further, we use
\begin{eqnarray}
  & \<k\la_e|\gamma^5 \hat S \gamma^\alpha |k' -\!\la_e\>
  \<P\la_p|\Gamma_\alpha|P'\la_p'\> = 
   4i\epsilon \left\{
  - M G_M \sh(\varphi(\la_p+\la_p')) \sin{\a/2}\,
  \sin\left({\a+\pi(\la_p'-\la_p)\over 2}\right) \right. + \nonumber\\
  & + {2M\over 4M^2+Q^2} (2\la_e)
    \left[ 2M\sqrt{s}G_E \ch(\varphi(\la_p-\la_p')) \cos{\a/2}\,
           \cos\left({\a+\pi(\la_p'-\la_p) \over 2}\right) +  \right.\\ 
  & \left. \left. +(s-M^2-Q^2/2) G_M \ch(\varphi(\la_p+\la_p')) \sin{\a/2}\,
            \sin\left({\a+\pi(\la_p'-\la_p) \over 2}\right) \right] \right\}\nonumber,
\end{eqnarray}
where $\sh \varphi = \epsilon/M$.

Putting all this together, we obtain 
\begin{eqnarray} \label{Bn}
  B_n = {4 m M \alpha \over D (4M^2+Q^2)} {(s-M^2)^2 \over s} (s-M^2-Q^2/2)
  \sin {\a/2}\, \ln^2{Q^2\over m^2}\, {\sum\limits_\n}' (2\pi)^3 \deltaPk
  \times \nonumber \\ \times
  \left\{ 2 \conj f_1 f_{-1} {\cal D}_{-1/2,3/2}
    \left( G_E {2M\sqrt{s} \over s-M^2-Q^2/2} \cos^2{\a/2} +
           G_M \sin^2{\a/2}\, {s+M^2 \over 2M\sqrt{s}} \right) + \right. \\
  \left. + {1\over 2} \eta_\n \sin \a 
     \left( |f_1|^2 {\cal D}_{3/2,-3/2} +
            |f_{-1}|^2 {\cal D}_{-1/2,1/2} \right)
     \left( G_E {s+M^2 \over s-M^2-Q^2/2} - G_M \right) \right\}, \nonumber
\end{eqnarray}
where ${\cal D}_{\la'\la} \equiv {\cal D}^{(s_\n)}_{\la'\la}(0,\a,0)$ and
$f_{\la} \equiv f^{(\n)}_{\la}(0)$.
Note that the summation is restricted to the hadronic states with a mass of $\sqrt{s}$ and thus (i) there is no elastic contribution to (\ref{Bn}) and (ii) at fixed $\a$ the dependence of $B_n$ versus energy has a resonant shape with maxima near existing baryon resonances.

\section{The limit of forward scattering\label{sec:Forward}}
In this section we compare our results to the analytic results of other authors.

Our expression (\ref{Bn}) obtained in the leading logarithm approximation 
is not restricted to small scattering angles
(the limits of our approximation will be discussed below).
This makes our result quite different from the results of \cite{AfanMer,Gorchtein}, where only forward kinematics was discussed.

Consider the forward scattering. Taking into account that
\begin{equation}
  {\cal D}_{\la'\la}(0,\a,0) \sim (\sin {\a/2})^{|\la-\la'|} \sim Q^{|\la-\la'|}
,\quad \text{at}\quad \a\to 0,
\end{equation}
we obtain the main contribution in the $\a \to 0$ limit:
\begin{eqnarray} \label{ourBn}
  B_n \approx {4 m M \alpha \over D (4M^2+Q^2)} {(s-M^2)^2 \over s}
  \sin {\a/2}\, \ln^2{Q^2\over m^2}\, {\sum\limits_\n}' (2\pi)^3 \deltaPk
  \times \nonumber \\ \times
  \left\{ 4M\sqrt{s} \conj f_1 f_{-1} {\cal D}_{-1/2,3/2} G_E +
  \eta_\n \sin {\a/2}\, |f_{-1}|^2 {\cal D}_{-1/2,1/2}
     \left( G_E (s+M^2) - G_M(s-M^2) \right) \right\}
\end{eqnarray}
with the scaling law
\begin{equation}
B_n \sim Q^3 \ln^2{Q^2\over m^2}\;.
\end{equation}
It agrees with the $Q$-dependence obtained in \cite{Gorchtein}.
However, the connection with the hadron transition amplitudes is different from the connection derived in \cite{Gorchtein}.
To compare our results, we have to adjust the notation. The helicity amplitudes of real Compton scattering $T_{\la'\la_p';\la\la_p}$, used in \cite{Gorchtein}, are related to the hadron photoproduction amplitudes by
\begin{equation}
  2 {\ \rm Im\ } T_{\la'\la_p';\la\la_p} =
  4\pi\alpha {\sum_\n}' (2\pi)^4 \deltaPk {\conj f}_{2\la_p'\la'} f_{2\la_p\la}
  \ \eta_\n^{\la_p-\la_p'}\ {\cal D}_{\la_p+\la,\la_p'+\la'}\;.
\end{equation}
Neglecting terms with higher powers of $\a$, the double-logarithmic contribution from Eq.~(23) of \cite{Gorchtein} reads
\begin{eqnarray} \label{gorBn}
B_n\approx  {4 m M \alpha \over D (4M^2+Q^2)}{(s-M^2)^2 \over s}
  \sin{\a/2}\, \ln^2{Q^2 \over m^2}\,
  {\sum\limits_\n}' (2\pi)^3 \deltaPk
  \times \nonumber \\ \times
  \left\{ 4M\sqrt{s} \conj f_1 f_{-1} {\cal D}_{-1/2,3/2} G_E +
         2M^2 \eta_\n \sin {\a/2}\, |f_{-1}|^2 {\cal D}_{-1/2,1/2} G_E
  \right\},
\end{eqnarray}
which is not identical to (\ref{ourBn}). Contrary to our result, the magnetic formfactor $G_M$ does not enter (\ref{gorBn}). Note, that this difference becomes crucial for the neutron target, because $G_E \approx 0$ for the neutron and the expression (\ref{gorBn}) just vanishes in that case.

In \cite{AfanMer} $B_n$ was studied using somewhat different approach.
While we picked up terms with the highest power of the large logarithm,
the authors of \cite{AfanMer} looked for the terms with the lowest
power of $Q$ in the limit of forward scattering. Their result behaves like:
\begin{equation} \label{AfBn}
 B_n \sim Q \ln {Q^2 \over m^2}.
\end{equation}
We see that the pre-logarithmic multiplier is $Q$ instead of our $Q^3$, but
on the other hand the logarithm is not squared. In our approach such term
comes from the second integral in the r.h.s. of (\ref{int}).
Comparing (\ref{AfBn}) and (\ref{ourBn}) one can figure out the applicability
conditions for each of the expressions.
First, both of them need
\begin{equation}
 \ln {Q^2 \over m^2} \gg 1.
\end{equation}
To be able to neglect (\ref{AfBn}) compared to (\ref{ourBn}), one also 
needs
%
\begin{equation} \label{app}
 \sin^2 \a/2 \, \ln {Q^2 \over m^2} \gg 1,
\end{equation}
under such condition our approximation is justified. On the other hand,
for the validity of (\ref{AfBn}), the inverse relation is needed,
$\sin^2 \a/2 \, \ln {Q^2 \over m^2} \ll 1$. In this case the scattering angle
should be extremely small, because of the large logarithmic factor in the l.h.s.
\section{Numerical results\label{sec:Numerical}}
We have calculated numerical values of $B_n$ for $e p \to e p$ using formula (\ref{Bn}), at electron lab. energies $E_{\rm lab}=$ 0.2, 0.3, 0.57 and 0.855 GeV. The first value corresponds to the SAMPLE experiment \cite{SAMPLE}, the second is near the Delta resonance peak and the last two are MAMI experimental points \cite{MAMI}.
Those energies are not far from the threshold, and thus the parameter
$\delta$, entering (\ref{int}), is not small. So, to obtain more accurate result
we also take into account last two terms in the square bracket in (\ref{int}).
This corresponds to replacing
\begin{equation}
 \ln^2 {Q^2 \over m^2} \to \ln^2 {Q^2 \over m^2} -
 8 \left[ F(\delta) + \ln (1+\delta) \ln {Q \over m \delta} \right]
\end{equation}
in the formula (\ref{Bn}).

We assume that the
intermediate hadronic states are three lightest resonances, $P_{33}(1232)$, $D_{13}(1520)$ and $S_{11}(1535)$. We also add the contribution from the threshold pion production in the $s$-wave, which is especially important for $E_{\rm lab} = 0.2$ GeV. In this case the intermediate hadronic state $\n$ is $\pi^+ n$ with the total angular momentum $s_\n = \half$ and $\eta_\n = -1$.

Thus we substitute
\begin{eqnarray}
 & & {\sum_\n}' (2\pi)^3 \deltaPk f^{(\n)}_\la(0) \conj f\!^{(\n)}_{\la'}(0) \to \\
 & & {4 W |\vec k_\pi| \over \pi \alpha} |E_{0+}(W)|^2 \delta_{\la,-1} \delta_{\la',-1} +
  \sum_R f^{(R)}_\la(0) \conj f\!^{(R)}_{\la'}(0) {\Gamma_R M_R \over \pi}
  {1 \over (W^2-M_R^2)^2 + M_R^2 \Gamma_R^2} \nonumber.
\end{eqnarray}
The first term comes from the threshold pion production,
$|\vec k_\pi| = {1 \over 2W} \sqrt{(W^2-M^2+m_\pi^2)^2-4 W^2 m_\pi^2}$ is the
pion c.m. momentum and $E_{0+}$ is the multipole amplitude.
The second term is the sum of the resonance contributions, $M_R$ and $\Gamma_R$
are the resonance mass and width, the quantities $f^{(R)}(0)$ are related to
$A_{3/2}$ and $A_{1/2}$, listed by PDG, as
\begin{equation}
f^{(R)}_1(0) = \sqrt{M(W^2-M^2) \over \pi\alpha} A_{3/2},\quad
  \eta_R f^{(R)}_{-1}(0) = \sqrt{M(W^2-M^2) \over \pi\alpha} A_{1/2}.
\end{equation}
All resonance parameters are from \cite{PDG}. The multipole amplitude $E_{0+}$ was taken from the MAID analysis \cite{MAID}. For the nucleon formfactors we use the well-known dipole fit.

\begin{figure}
  \includegraphics[height=0.2\textheight]{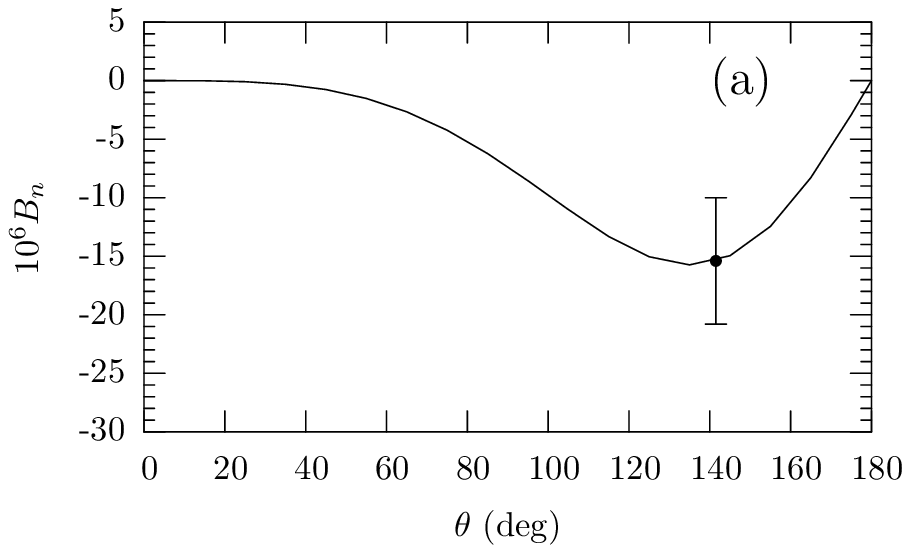}
  \includegraphics[height=0.2\textheight]{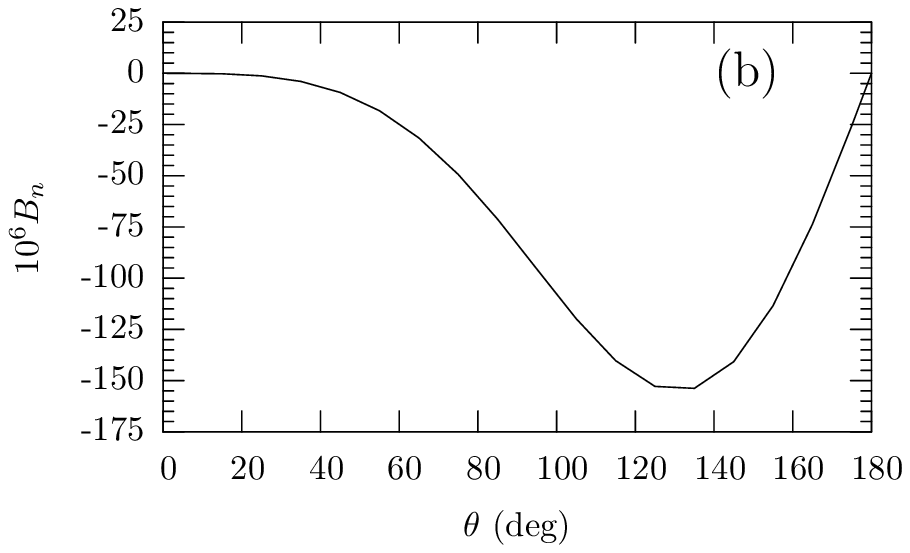}
  \includegraphics[height=0.2\textheight]{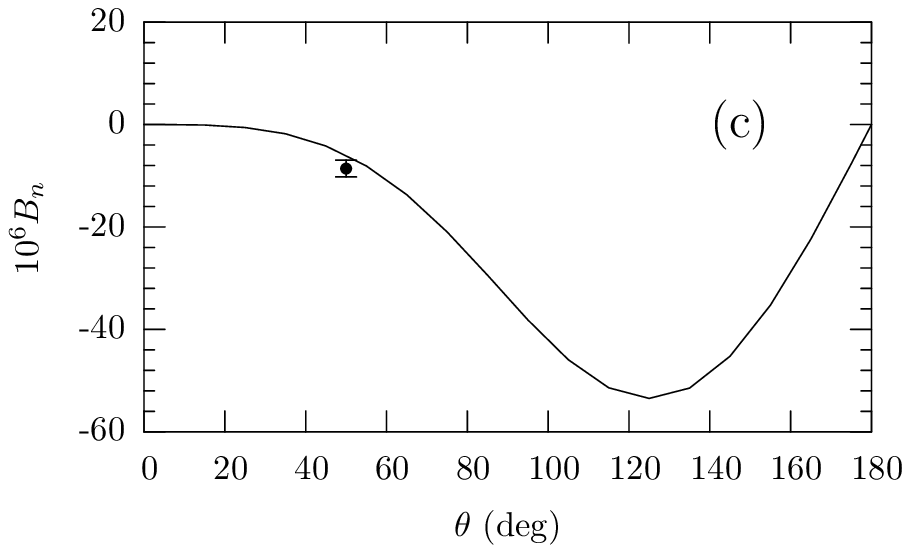}
  \includegraphics[height=0.2\textheight]{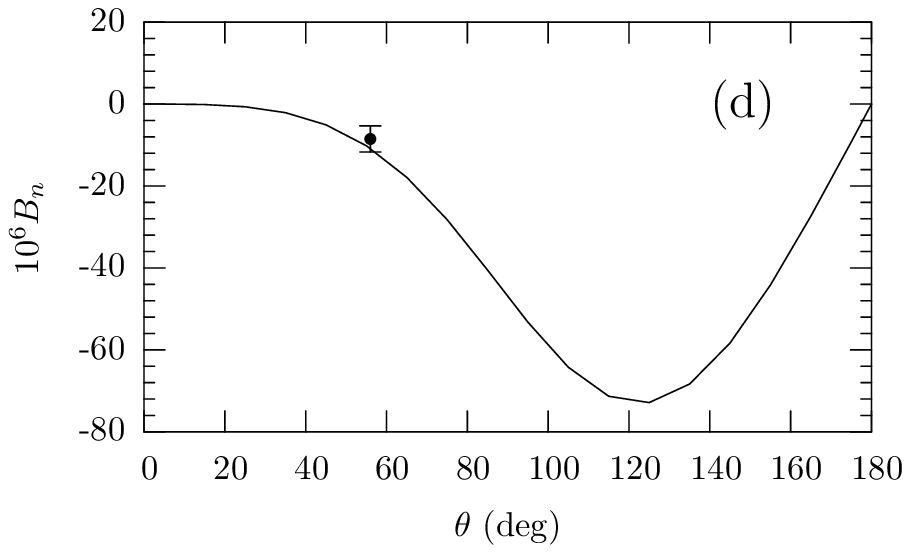}
  \\
 \caption{\label{fig.proton} The beam normal spin asymmetry of the elastic $ep$ scattering at electron energy (a) 0.2 GeV, (b) 0.3 GeV, (c) 0.57 GeV and (d) 0.855 GeV. Experimental points are from \cite{SAMPLE,MAMI}.} 
\end{figure}

The dependence of $B_n$ versus the c.m. scattering angle is displayed on Figure~\ref{fig.proton} together with the experimental data. One can see that the calculated asymmetry for the proton target agrees with experiment.

We also compare our results with the numerical calculation of Ref.~\cite{Pasq}, where $N$ and $\pi N$ intermediate hadronic states were taken into account.
At $E_{\rm lab}=0.2$~GeV Ref.~\cite{Pasq} gives the result, quite different from our one, as well as from the experiment. At higher energy the shape of the curves is similar, nevertheless we predict the maximal absolute value about 1.5 - 3 times larger than \cite{Pasq}. At present we cannot find an explanation for this discrepancy.

\begin{figure}
  \includegraphics[height=0.2\textheight]{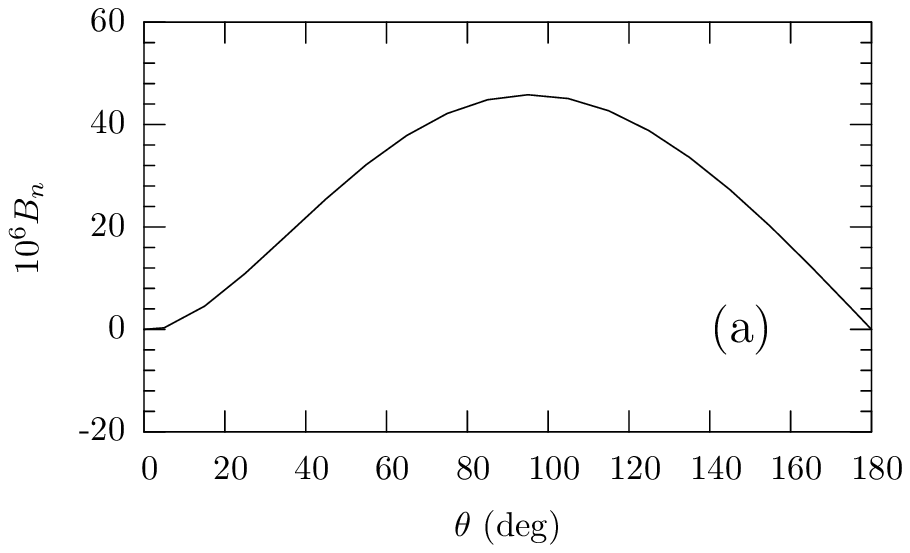}
  \includegraphics[height=0.2\textheight]{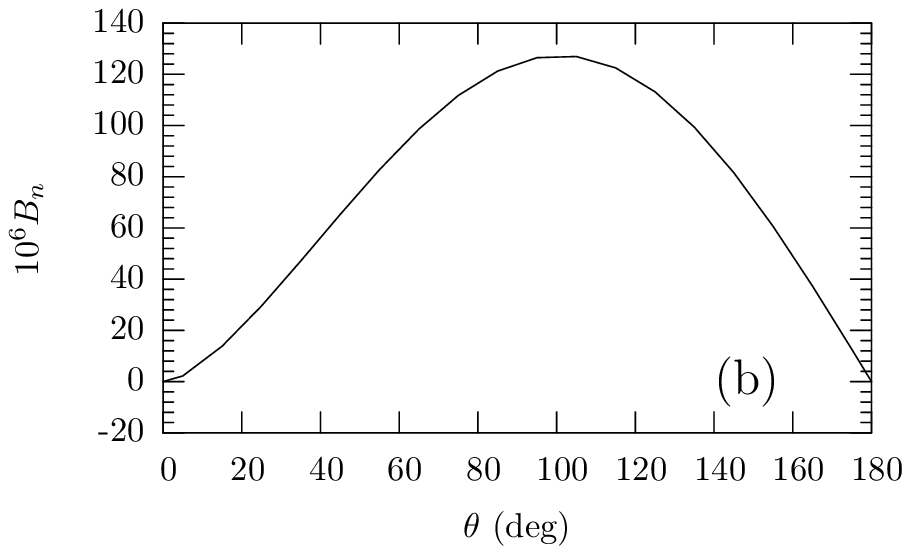}
  \includegraphics[height=0.2\textheight]{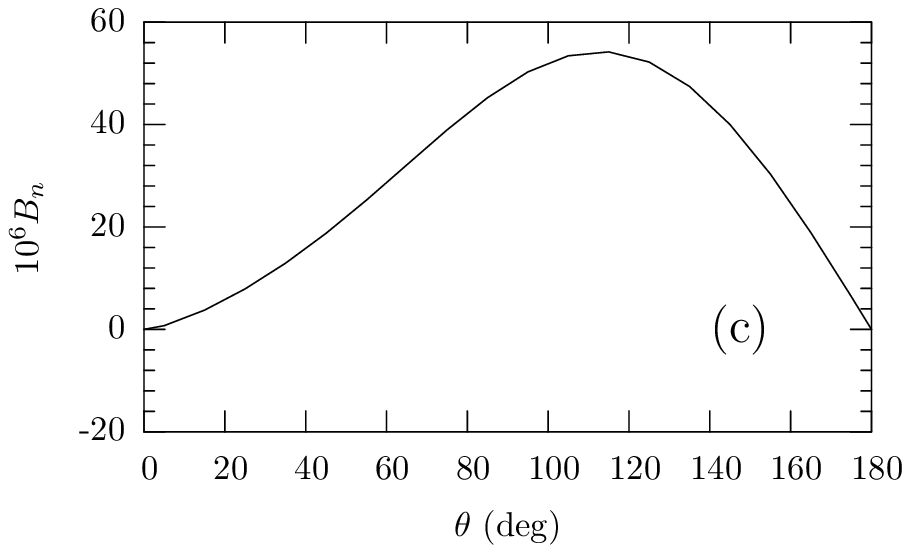}
  \includegraphics[height=0.2\textheight]{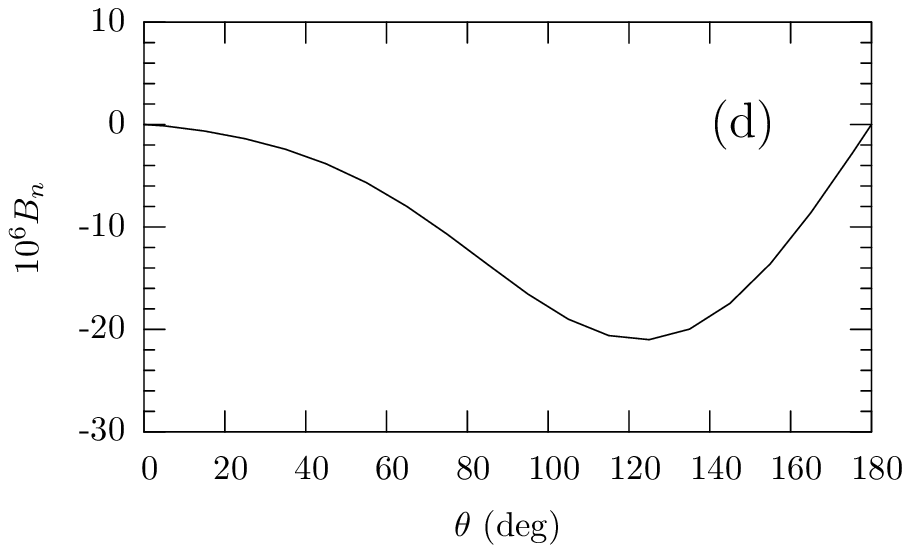}
  \\
 \caption{\label{fig.neutron} The beam normal spin asymmetry of the elastic $en$ scattering at electron energy (a) 0.2 GeV, (b) 0.3 GeV, (c) 0.57 GeV and (d) 0.855 GeV.} 
\end{figure}

The asymmetry for the neutron target can be calculated in the same way as for the proton. The results are given on Figure~\ref{fig.neutron}.
\section{Conclusion\label{sec:Conclusions}}
We have presented  analysis of the beam normal spin asymmetry $B_n$ of the elastic $eN$ scattering in the leading logarithm approximation.

Contrary to results of other authors \cite{AfanMer,Gorchtein} obtained for the forward kinematics only, our results are valid for any scattering angles, which fulfill the condition
$\sin^2 \a/2 \, \ln {Q^2 \over m^2} \gg 1$. The results of \cite{AfanMer}
are valid if the inverse relation is satisfied.

For the forward kinematics we get the following $Q$-dependence: $B_n \sim Q^3 \ln^2(Q^2/m^2)$, which coincides with prediction of \cite{Gorchtein}. Nevertheless we get the relation between $B_n$ and the hadron electromagnetic transition amplitudes, which differs from \cite{Gorchtein}.

We calculate numerical values of $B_n$ on the proton target at intermediate energy, and show that the results are consistent with existing experimental data. We also give a prediction for the asymmetry on the neutron target. 

\appendix
\section{Integral calculation}
In this section we denote $k \equiv |\vec k|$, $k'' \equiv |\vec k''|$; this
cannot lead to a confusion with 4-momenta, since they are not used hereafter.

The integral to calculate is
\begin{equation}
  \int {d^3 k'' \over 2\epsilon''}{1 \over q_1^2 q_2^2} Y =
  I_1 + I_2 + I_3,
\end{equation}
where
\begin{eqnarray}
  I_1 & = & \int {d^3 k'' \over 2\epsilon''}{1 \over q_1^2 q_2^2} \widetilde Y, \nonumber \\
  I_2 & = & \int {k''^2 dk'' \over 2 \epsilon''} [Y_0(W)-Y_0(\Wmax)] \int {d\Omega'' \over q_1^2q_2^2}, \\
  I_3 & = & Y_0(\Wmax) \int {k''^2 dk'' \over 2 \epsilon''} \int {d\Omega'' \over q_1^2q_2^2} \nonumber\; ,
\end{eqnarray}
where $\widetilde Y = Y - Y_0$,
$Y_0(W) = Y(W,0,-4 k k'' \sin^2 {\a/2})$.
The function $\widetilde Y$ is zero at $q_1^2=0$ and at $q_2^2=0$
by construction, thus the first integral, $I_1$, does not contain logarithmic terms.
The angular integration in the last two ones can be done using
\begin{equation}
\begin{split}
  \int {d\Omega'' \over q_1^2 q_2^2} & =  
  {1 \over 4k^2k''^2} \int {d\Omega'' \over (a\!-\!\cos\a_1)(a\!-\!\cos\a_2)} = \\ & = {1 \over 4k^2k''^2} {4\pi \over \sin {\a/2} \sqrt{a^2\!-\!\cos^2{\a/2}}}
  \ln {\sin{\a/2}+\sqrt{a^2\!-\!\cos^2{\a/2}} \over \sqrt{a^2-1}},
\end{split}
\end{equation}
where $a = {\epsilon\epsilon'' - m^2 \over k k''}$. The quantity $a$ varies
from $a_{\rm min} = \sqrt{1+(m\delta/k)^2}$ to $\infty$. However $a\approx 1$
for almost all $k''$ values and can be substantially greater only for $k'' \sim m$.
Since $Y_0(W)-Y_0(\Wmax)=0$ at $k''=0$, we may put $a=1$ in the integral $I_2$
everywhere besides the denominator under the logarithm; one can prove that
\begin{eqnarray}
 I_2 = {\pi \over 2 k^2 \sin \a/2}
 \int {dk'' \over \epsilon''} [Y_0(W)-Y_0(\Wmax)]
 {1 \over \sqrt{a^2-\cos^2{\a/2}}}
 \ln {\sin{\a/2}+\sqrt{a^2-\cos^2{\a/2}} \over \sqrt{a^2-1}} = \nonumber \\
 = \frac{\pi}{2k^2\sin^2{\theta/2}} \int {dk'' \over \epsilon''} [Y_0(W)-Y_0(\Wmax)] \ln {2\sin{\a/2}\over \sqrt{a^2-1}}+\mathcal{O}(m) = \\ 
 = - {2\pi \over Q^2} \int_{\Wth^2}^s dW^2
 {Y_0(W)-Y_0(\sqrt{s}) \over W^2-s} \ln {Q \over m}{s-W^2 \over W^2-M^2}
 + \mathcal{O} (m). \nonumber 
\end{eqnarray}
The integral $I_3$ is
\begin{equation}
 I_3 = {\pi Y_0(\Wmax) \over 2k^2 \sin \a/2}
 \int {dk'' \over \epsilon''} {1 \over \sqrt{a^2-\cos^2{\a/2}}}
 \ln {\sin{\a/2} + \sqrt{a^2-\cos^2{\a/2}} \over \sqrt{a^2-1}}.  
\end{equation}
We introduce new independent variable $t$ defined according to
$a^2 = 1 + {\sin^2{\a/2} \over \sh^2 t}$,
$0 \leq t \leq {\rm Arsh\ } {k\sin \a/2 \over m\delta}$.
In the following we denote $\sigma \equiv \sin \a/2$.
After that the integral becomes
\begin{equation}
\begin{split}
 I_3 & = {\pi Y_0(\Wmax) \over 2k\sigma} \int {t \sh t\ dt \over \sqrt{\sigma^2 + \sh^2 t}
 (\epsilon\sigma + m \sqrt{\sigma^2 + \sh^2 t})} = \\
 & = {\pi Y_0(\Wmax) \over 2 \epsilon k \sigma^2} \left[
 \int \left({\sh t \over \sqrt{\sigma^2 + \sh^2 t}} - 1 \right) t dt +
 \int \left(1 - {\sh t \over \epsilon\sigma/m + \sqrt{\sigma^2 + \sh^2 t}}
      \right) t dt
 \right].
\end{split}
\end{equation}
The first integral in the square bracket after substitution $t={1\over 2} \ln {u-u^2 \over u-\sigma^2}$, $\sigma^2 \leq u \leq \sigma$, reduces to
\begin{equation}
  {1\over 2} \int {du \over 1-u} \ln {u-u^2 \over u-\sigma^2} = 
  -{1\over 4} \ln^2 {1-u \over 1-\sigma^2} - {1\over 2} F(u-1) +
  {1\over 2} F \left({u-1 \over 1-\sigma^2} \right),
\end{equation}
where $F$ is Spence function, defined by (\ref{Spence}).
The second integral, up to $\mathcal{O}(m)$ terms, can be rewritten as
\begin{equation}
  \int \left(1 - {\sh t \over \epsilon\sigma/m + \sqrt{\sigma^2 + \sh^2 t}}
      \right) t dt \approx 
  \int \left(1 - {e^t \over 2\epsilon\sigma/m + e^t} \right) t dt = 
  {1\over 2} \ln^2 \left(e^{-t}+{m\over 2\epsilon \sigma} \right) -
  F \left({-1 \over {2\epsilon\sigma \over m} e^{-t} + 1} \right).
\end{equation}
The intuitive reason is that the quantity $\epsilon \sigma \over m$ in the
denominator is large, so $\sh t$ should also be large for the fraction
to be non-zero, thus we may put $\sh t \approx e^t/2$.

Inserting the integration limits, we obtain the final result
\begin{equation}
 I_3 = {2\pi \over Q^2} \left[ 
  {1 \over 8} \ln^2 {Q^2 \over m^2} + {\pi^2 \over 8}
  - {1 \over 4} F(-\cos^2 {\a/2})
  - F(\delta) - \ln (1+\delta) \ln {Q \over m \delta} \right].
\end{equation}

\def\Jou#1#2#3#4#5{#1, #2 {\bf #3} (#4), #5}

\end{document}